\documentclass[a4paper,11pt]{article}
\usepackage{url}
\usepackage{pos}
\usepackage{caption}
\usepackage{subcaption}

\title{An efficient hit finding algorithm for Baikal-GVD muon reconstruction}

\author[a]{V.A.~Allakhverdyan}
\author*[b]{A.D.~Avrorin}
\author[b]{A.V.~Avrorin}
\author[b]{V.M.~Aynutdinov}
\author[c]{R.~Bannasch}
\author[d]{Z.~Barda\v{c}ov\'{a}}
\author[a]{I.A.~Belolaptikov}
\author[a]{I.V.~Borina}
\author[a,1]{V.B.~Brudanin}
\author[e]{N.M.~Budnev}
\author[a]{V.Y.~Dik}
\author[b]{G.V.~Domogatsky}
\author[b]{A.A.~Doroshenko}
\author[a,d]{R.~Dvornick\'{y}}
\author[e]{A.N.~Dyachok}
\author[b]{Zh.-A.M.~Dzhilkibaev}
\author[d]{E.~Eckerov\'{a}}
\author[a]{T.V.~Elzhov}
\author[f]{L.~Fajt}
\author[g,1]{S.V.~Fialkovski}
\author[e]{A.R.~Gafarov}
\author[b]{K.V.~Golubkov}
\author[a]{N.S.~Gorshkov}
\author[e]{T.I.~Gress}
\author[a]{M.S.~Katulin}
\author[c]{K.G.~Kebkal}
\author[c]{O.G.~Kebkal}
\author[a]{E.V.~Khramov}
\author[a]{M.M.~Kolbin}
\author[a]{K.V.~Konischev}
\author[h]{K.A.~Kopa\'{n}ski}
\author[a]{A.V.~Korobchenko}
\author[b]{A.P.~Koshechkin}
\author[i]{V.A.~Kozhin}
\author[a]{M.V.~Kruglov}
\author[b]{M.K.~Kryukov}
\author[g]{V.F.~Kulepov}
\author[h]{Pa.~Malecki}
\author[a]{Y.M.~Malyshkin}
\author[b]{M.B.~Milenin}
\author[e]{R.R.~Mirgazov}
\author[a]{D.V.~Naumov}
\author[a]{V.~Nazari}
\author[h]{W.~Noga}
\author[b]{D.P.~Petukhov}
\author[a]{E.N.~Pliskovsky}
\author[j]{M.I.~Rozanov}
\author[a]{V.D.~Rushay}
\author[e]{E.V.~Ryabov}
\author[b]{G.B.~Safronov}
\author*[a]{B.A.~Shaybonov}
\author[b]{M.D.~Shelepov}
\author[a,d,f]{F.~\v{S}imkovic}
\author[a]{A.E. Sirenko}
\author[i]{A.V.~Skurikhin}
\author[a]{A.G.~Solovjev}
\author[a]{M.N.~Sorokovikov}
\author[f]{I.~\v{S}tekl}
\author[b]{A.P.~Stromakov}
\author[a]{E.O.~Sushenok}
\author[b]{O.V.~Suvorova}
\author[e]{V.A.~Tabolenko}
\author[e]{B.A.~Tarashansky}
\author[a]{Y.V.~Yablokova}
\author[c]{S.A.~Yakovlev}
\author[b]{D.N.~Zaborov}

\affiliation[a]{Joint Institute for Nuclear Research, Dubna, Russia}
\affiliation[b]{Institute for Nuclear Research, Russian Academy of Sciences, Moscow, Russia}
\affiliation[c]{EvoLogics GmbH, Berlin, Germany}
\affiliation[d]{Comenius University, Bratislava, Slovakia}
\affiliation[e]{Irkutsk State University, Irkutsk, Russia}
\affiliation[f]{Czech Technical University in Prague, Prague, Czech Republic}
\affiliation[g]{Nizhny Novgorod State Technical University, Nizhny Novgorod, Russia}
\affiliation[h]{Institute of Nuclear Physics of Polish Academy of Sciences (IFJ~PAN), Krak\'{o}w, Poland}
\affiliation[i]{Skobeltsyn Institute of Nuclear Physics, Moscow State University, Moscow, Russia}
\affiliation[j]{St.~Petersburg State Marine Technical University, St.Petersburg, Russia}

\note{Deceased.}

\emailAdd{avrorin@inr.ru}
\emailAdd{bairsh@yandex.ru}

\abstract{
  The Baikal-GVD is a large scale neutrino telescope being constructed in Lake Baikal.
  The majority of signal detected by the telescope are noise hits, caused primarily by the luminescence of the Baikal water.
  Separating noise hits from the hits produced by Cherenkov light emitted from the muon track is a challenging part of the muon event reconstruction.
  We present an algorithm that utilizes a known directional hit causality criterion to contruct a graph of hits and then use a clique-based technique to select the subset of
  signal hits.
  The algorithm was tested on realistic detector Monte-Carlo simulation for a wide range of muon energies and has proved to select a pure sample of PMT hits from Cherenkov
  photons while retaining above 90\% of original signal.
}

\FullConference{%
  *** 37th International Cosmic Ray Conference (ICRC2021), ***\\
  *** 12-23 July 2021 ***\\
  *** Berlin, Germany - Online ***
}

\begin{document}

\maketitle

\section{Introduction}
Baikal-GVD \cite{gvd, gvd2021} is a cubic kilometer scale underwater neutrino telescope currently under construction in the southern basin of Lake Baikal.
The telescope operates by detecting Cherenkov radiation from neutrino induced secondary particles in the Baikal water with an array of PMTs submerged between the depths of 750 and 1275 meters.
The PMTs are organized into clusters, with 288 PMTs in each cluster.
In its present configuration Baikal-GVD consists of 8 clusters and is currently taking data.

Most of the PMT hits are produced by noise rather than Cherenkov photons.
While there are several sources of noise in GVD, including PMT dark current and electronic noise, they are dominated by 1 p.e. signal from the Baikal water luminiscence \cite{noise2019icrc}.
The PMT noise counting rates can vary between 10 and 200 kHz and, as can be seen in \cite{lum2021}, exhibit heavy variation with time and elevation.
Suppressing noise hits while preserving signal is of particular importance for the muon track reconstruction where a charge deposition on a single PMT can often be comparable to the noise.

We present an approach for selecting signal hits based on the ScanFit technique introduced in \cite{bruijn2008antares}. 
Similarly to ScanFit, for each event we assume a single muon track model and cover the sky with the lattice of possible track origin points corresponding to a considered track direction.
A set of hits is selected for each direction using a directional causality test (called 1-D clustering in \cite{bruijn2008antares}) and the set that produces the optimal hit time based fit for a single track hypothesis is considred to be the signal, providing both noise suppression and a track prefit for reconstruction.

Our approaches to hit selection for a particular direction and picking the best direction, however, are novel.
For each considered direction we are using a maximal clique search algorithm to find subsets of hits where each pair of hits passes a directional causality test. 
The hits in the largest or second-largest subset with an optimal signal track fit are then selected as the signal.
The direction of the winning fit can also be used as a track prefit for track reconstruction.
Due to the computational cost of clique search for multiple directions we also apply a prefilter to remove easily identifiable noise hits while keeping as much of the signal as possible.

This approach operates on the assumption that if we treat the directional causality condition as a binary test for a pair of hits, each pair of signal hits should pass, while a noise hit is unlikely to pass it in a pair with every signal hit in the event. 
This prevents rejecting otherwise isolated hits, increasing hit selection efficiency and, as the number of signal hits increases, the number of causality tests a noise hit must pass grows as well, increasing hit selection purity.
For an atmospheric neutrino sample and low noise conditions, the resulting hit selection has an average hit-wise purity and efficiency of 95\%.

\section{Directional causality}
In this work we heavily use a directional causality criterion from \cite{bruijn2008antares}. 
For a predefined track direction and a pair of causaly related hits $i$ and $j$, the following conditions must be true:
\begin{align*}
  | \Delta r_{i,j} | < R
  \\
  \Delta z_{i,j} - kR - w \leq c \Delta t_{i,j} \leq \Delta z_{i,j} + kR + w
\end{align*}

Here, $\Delta r_{i,j}$ is the difference between hit PMTs distances to the muon track, $\Delta z_{i,j}$ is the distance between hit PMTs along the muon track, $R$ is the distance between hit PMTs in the plane orthogonal to the track, $\Delta t_{i,j}$ is time difference between hits, $w$ is a tunable parameter used to compensate for hit timing uncertainty and $k$ is a constant depending only on the refraction index of the observed medium.
Because the track direction is predefined, the time window imposed by this criterion on a pair of hits is narrow and noise hits are less likely to fall into it. 

Note that the directional causality relies only track direction, which makes the scanning part of the primary algorithm possible.

\section{Algorithm description}
The hit selection algorithm is performed in three steps: initial noise rejection, scanning the sky and selecting hit subsets for each direction, and selecting the optimal direction and a subset of hits. 
Each step is described in a subsection below.

\subsection{Prefilter}
\label{subs:prefilter}
At this stage a set of hits is preselected for processing.
Processing time in the following stages increases quadratically with the number of hits and linearly with the number of considered directions, so at this stage we want to remove easily identifiable noise.
This is accomplished in three steps.
First, we construct a causality graph of the event - a graph where each hit is represented by a vertex and two vertices share an edge if the corresponding hits satisfy causality conditions as described in \cite{allakhverdyan2021measuring}.
The causality parameters are relaxed to avoid rejecting signal hits.

Next, we use the Bron-Kerbosch algorithm \cite{bron1973algorithm} to find the maximal cliques in the causality graph - the largest subsets of vertices where all vertices are connected to each other.
This way, each clique corresponds to a subset of hits where each pair of hits satisfies the causality conditions.
Once the maximal cliques are found, all the hits that are not in the largest clique are rejected. 
If there are multiple largest cliques of equal size, the hits from all of these cliques are preserved.

Finally, the events with fewer then 6 hits on 3 strings are discarded. 
This primarily filters purely noise induced events.
For an atmospheric neutrino sample and low noise conditions, average prefilter hit-wise efficiency is 98\% and average hit-wise purity is 69\%.

\subsection{Scan}
A set of fixed directions in the sky - a scan grid - is considered.
In this work we use a rectangular grid in spherical coordinates as a scan grid.
Each direction in the scan grid is a muon origin point and is associated with a muon track parametrized by the tuple $(\theta, \phi, X_0, Y_0, T_0)$, where $\theta$ and $\phi$ are polar and azimuth angle of the track, $X_0$ and $Y_0$ are coordinates of the track projection to the plane orthogonal to the track direction and $T_0$ is the time when the track crosses this plane. 
For each track $\theta$ and $\phi$ are fixed, while $X_0$, $Y_0$ and $T_0$ are free parameters.

For each direction in the scan grid, similarly to the prefilter step, we construct a causality graph from the event hits.
This time, however, we are using directional causality to establish whether two vertices corresponding to a pair of hits should share an edge.
Because each direction in the scan grid is associated with a different track model, each direction will have its own causality graph.

Next, we use Bron-Kerbosch search to find maximal cliques for all the graphs considered at the previous step.
Here, minimum clique size is a tunable parameter and directions without maximal cliques above minimum clique size are rejected from further consideration.
Minimum clique size is set to the minimum number of hits required for reconstruction and typically equals 8 hits.
For each direction we then pick the largest cliques in the graph associated with it, called the maximum cliques.
A direction may have multiple maximum cliques, but they all should be of the same size.
All cliques beside the maximum cliques are rejected from further consideration.

In this approach we are using the Bron-Kerbosch search that produces all maximal cliques in the graph rather then a faster equivalent that would only search for maximum cliques.
The reasoning behind this decision is that the number of maximal cliques associated with a direction can inform the assumptions about topology of the event (such as the number of muons and showers) that may be useful for further steps in the muon track reconstruction pipeline.

\begin{figure}
  \centering
  \includegraphics[height=5.5cm]{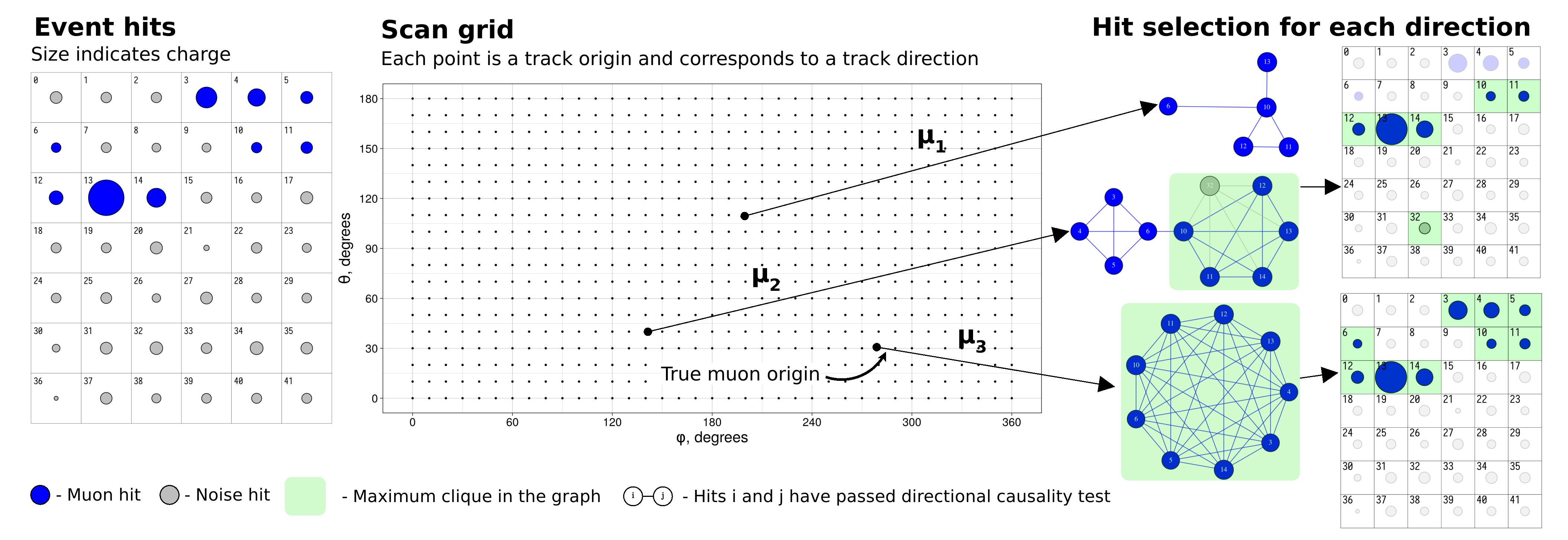}
  \caption{Example of the scanning step for an upgoing muon event.}
  \label{fig:scan_step_example}
\end{figure}

At the end of this step, each scan grid direction should be associated with a muon track, a causality graph and a set of cliques of equal size.
The example of the scanning step for 3 considered track directions of an upgoing muon event with the minimal clique size set to 6 is presented on Figure \ref{fig:scan_step_example}.
Three scan grid directions are considered. 
$\mu_1$ produces a causality graph with the maximum clique of size 3 and is ignored.  
$\mu_2$ produces a graph with two maximal cliques and only the largest one is considered. 
$\mu_3$ is close to the true track direction and produces a graph with only one clique that contains all signal hits.
Finally, two cliques pass to the clique selection stage: a single clique from $\mu_3$ and the largest clique from $\mu_2$.

\subsection{Clique selection}

Once the maximum cliques for each direction in the scan grid are found, we determine the largest clique size for the event $N_{max}$ and reject all cliques with sizes below $N_{max} - 1$.
This step typically rejects the majority of directions in the grid, as they would no longer have a single clique associated with them.
For each of the remaining cliques, we fit the associated tracks on the space of $X_0$, $Y_0$ and $T_0$ to a single track model. This is accomplished with a hit time based M-estimator, to mitigate effect of the outliers:
\begin{align*}
  M = \frac{1}{N - 2}\sum_i^{N}{ln(1 + \frac{(T_0 + t^{th}_i(X_0, Y_0) - t_i)^2 }{2\sigma^2})}
\end{align*}
If the M-estimator minimization does not converge, the corresponding clique is rejected.

Following this step, each remaining clique is associated with a fully-defined track and its fit quality represented by the M-estimator score.
Finally, we select the clique with the best M-estimator value.
The hits of this clique are considered to be the signal hits of the event and the associated track can be used as a prefit for a more comprehensive reconstruction technique.

\section{Performance and results}
\begin{figure}
     \centering
     \begin{subfigure}[t]{0.3\textwidth}
         \centering
         \includegraphics[width=\textwidth]{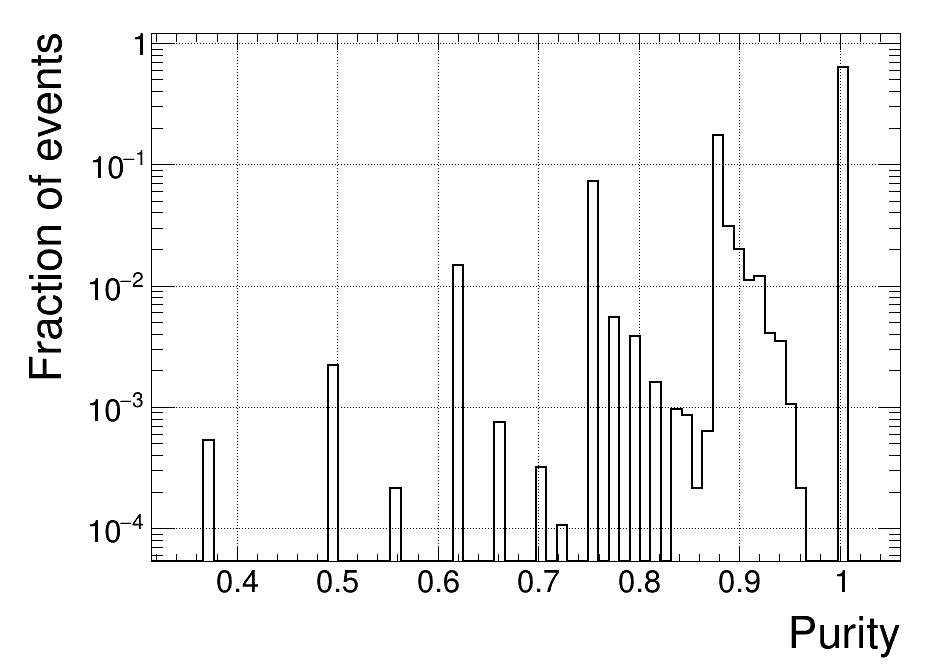}
         \includegraphics[width=\textwidth]{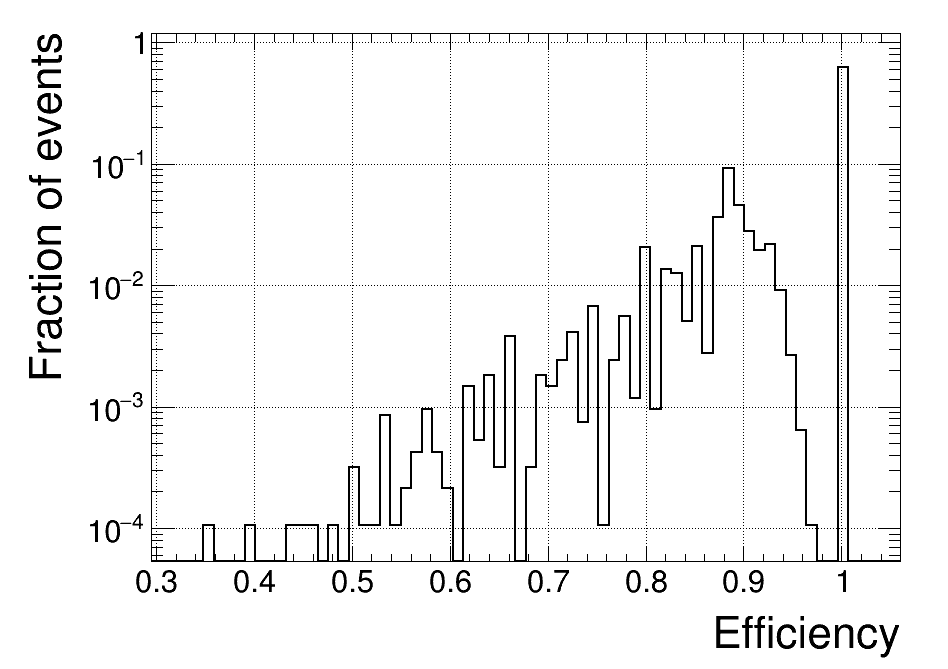}
         \caption{Atmospheric neutrino sample.}
         \label{fig:atmnu_results}
     \end{subfigure}
     \hfill
     \begin{subfigure}[t]{0.3\textwidth}
         \centering
         \includegraphics[width=\textwidth]{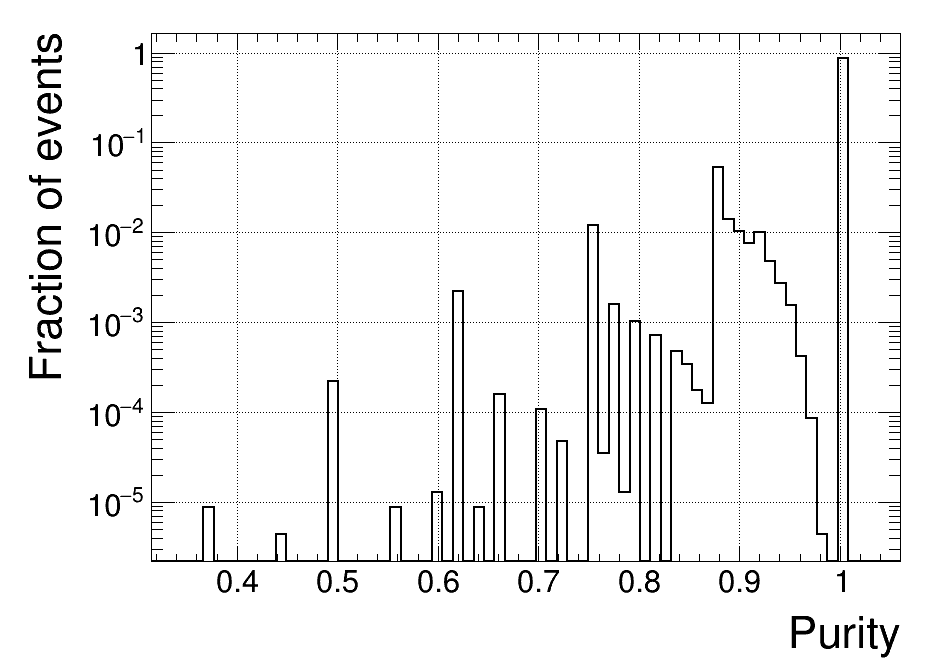}
         \includegraphics[width=\textwidth]{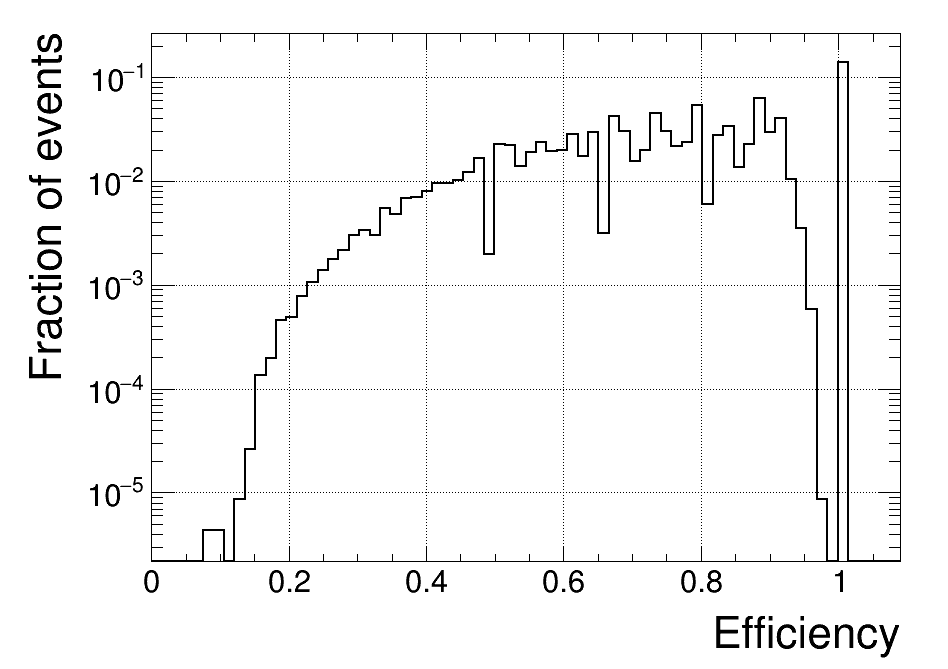}
         \caption{High energy ($E^{-2}$) neutrino sample.}
         \label{fig:nu2_results}
     \end{subfigure}
     \hfill
     \begin{subfigure}[t]{0.3\textwidth}
         \centering
         \includegraphics[width=\textwidth]{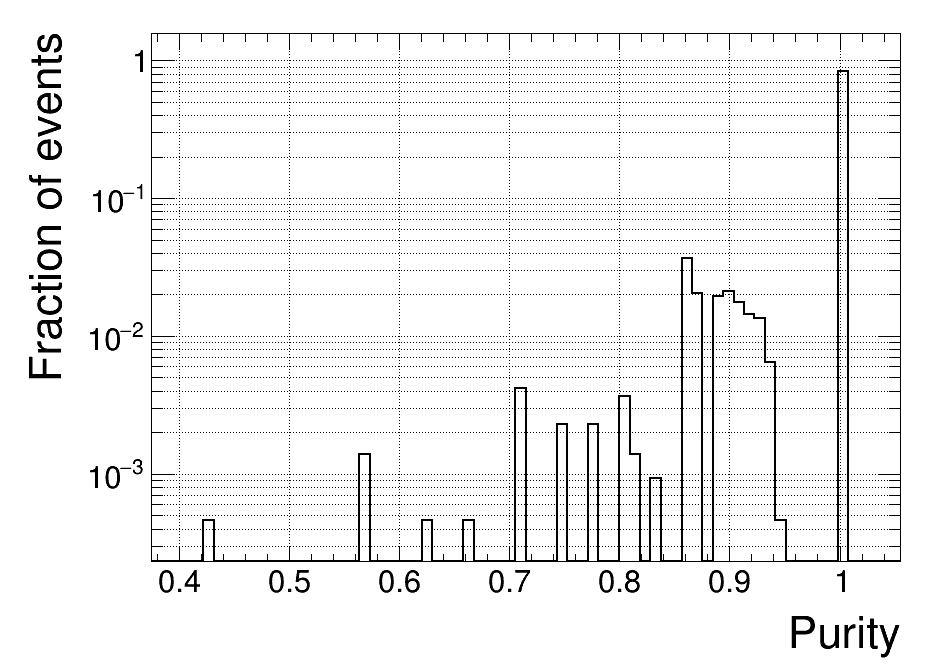}
         \includegraphics[width=\textwidth]{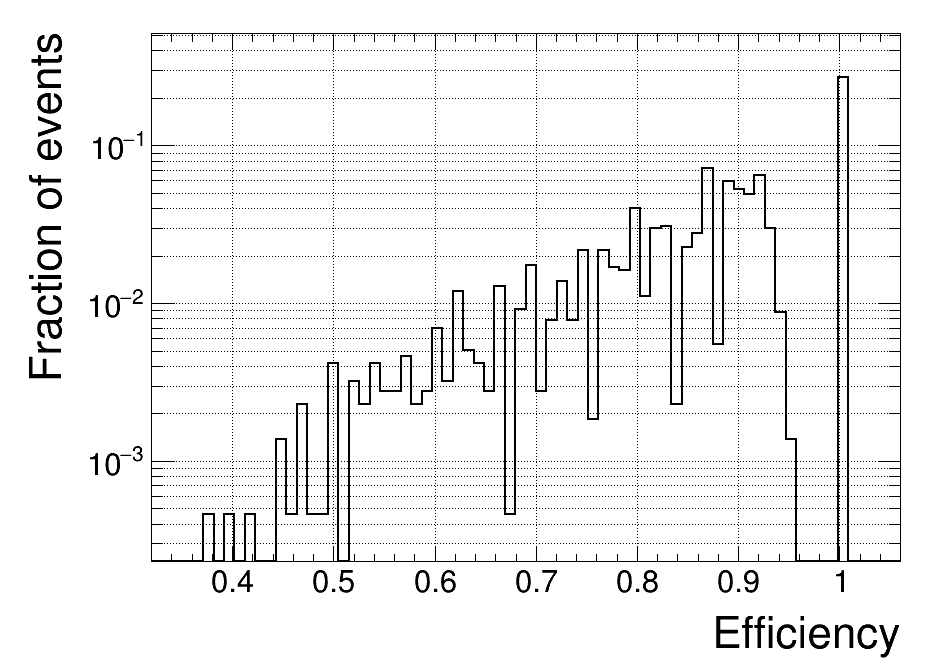}
         \caption{Multicluster atmospheric neutrino sample}
         \label{fig:numc_results}
     \end{subfigure}
     \hfill

     \caption{Hit-wise purity and efficiency distributions for various samples.}
     \label{fig:results}
\end{figure}

The performance of the presented appoach was studied on three samples. 
The first sample consists of single cluster atmospheric muon neutrinos, simulated with Bartol flux and the low threshold at 20 GeV.
The second sample is equivalent to the first one, except it was simulated with $E^{-2}$ spectrum.
Finally, the third sample is equivalent to the first one, however it only contains multicluster events.
All samples correspond to a low noise period in season 2019 characterized by PMT noise counting rates of $\sim 20$ kHz for all elevations. 
The samples have been processed with a rectangular scan grid with a step of 3 degrees in both $\theta$ and $\phi$, minimum clique size of 8 hits, and a directional causality window set to 20 ns.

The key estimated hit selection quality characterics for each sample in this performance study are purity and efficiency.
The purity of hit selection for an event (also known as precision, or positive predictive value) is defined as the ratio of the number of correctly identified signal hits to the total number of hits identified as a signal.
The efficiency of hit selection for an event (also known as recall, or true positive rate) is defined as the ratio of the number of correctly identified signal hits to the total number of signal hits in the event.
The distributions of purity and efficiency for the processed samples are presented on Figure \ref{fig:results}.

The average values of purity and efficiency for each sample are presented in Table \ref{tbl:results}.
Note that compared to the hit selection procedure used in \cite{allakhverdyan2021measuring}, the efficiency for atmospheric neutrino events has almost doubled.
Lower efficiency for the high energy and multicluster samples is explained by the increased fraction of the delayed cascade hits due to the higher energies of observed particles.
The directional causality uses the model of a single muon and so some cascade hits are excluded from cliques, decreasing overall efficiency of hit selection.
Importantly, cascade hits that fit the track model within the timing uncertainty can be included in cliques along  with the muon hits.
However, if we restrict the samples to events with at least 8 track hits, efficiency improves.

\begin{table}[h]
  \centering
  \caption{Hit selection quality characterics for various samples}
  \begin{tabular}{lll}
  \textbf{Sample}                    & \textbf{Average purity} & \textbf{Average efficiency} \\
    Atmospheric neutrino sample (hit selection from \cite{allakhverdyan2021measuring})              & 0.99                    & 0.5                         \\
    Atmospheric neutrino sample              & 0.95                    & 0.95                        \\
    High energy ($E^{-2}$) neutrino sample  & 0.98                    & 0.74                        \\
    High energy ($E^{-2}$) neutrino sample (at least 8 track hits)  & 0.98                    & 0.89                        \\
    Multicluster atmospheric neutrino sample & 0.98                  & 0.87 \\                       
    Multicluster atmospheric neutrino sample (at least 8 track hits) & 0.98                    & 0.9                       
  \end{tabular}
  \label{tbl:results}
\end{table}

\begin{figure}[b]
     \centering
     \begin{subfigure}[b]{0.47\textwidth}
         \centering
         \includegraphics[width=\textwidth]{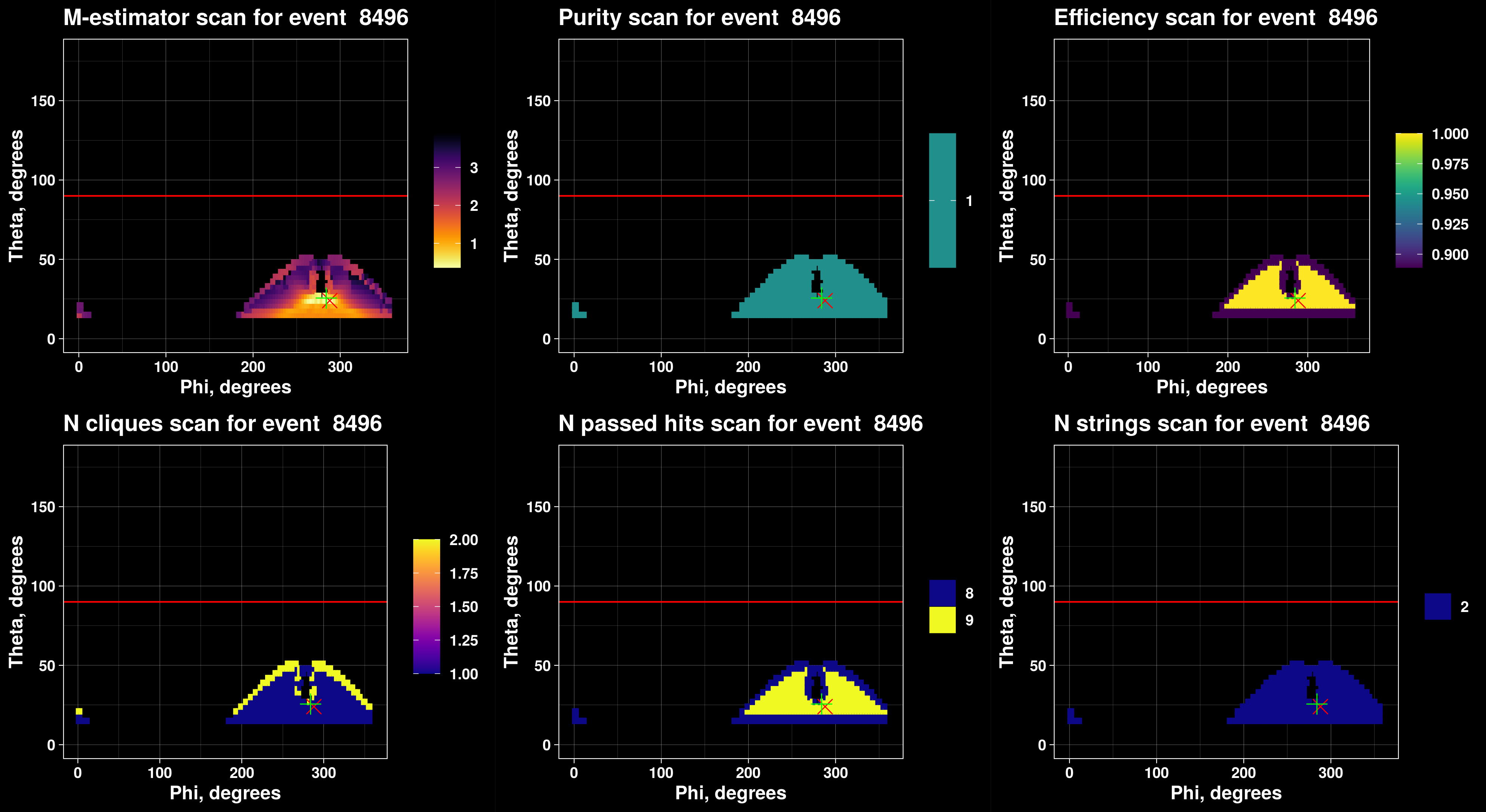}
         \caption{Upgoing muon from atmospheric neutrino}
         \label{fig:scan_up}
     \end{subfigure}
     \hfill
     \begin{subfigure}[b]{0.47\textwidth}
         \centering
         \includegraphics[width=\textwidth]{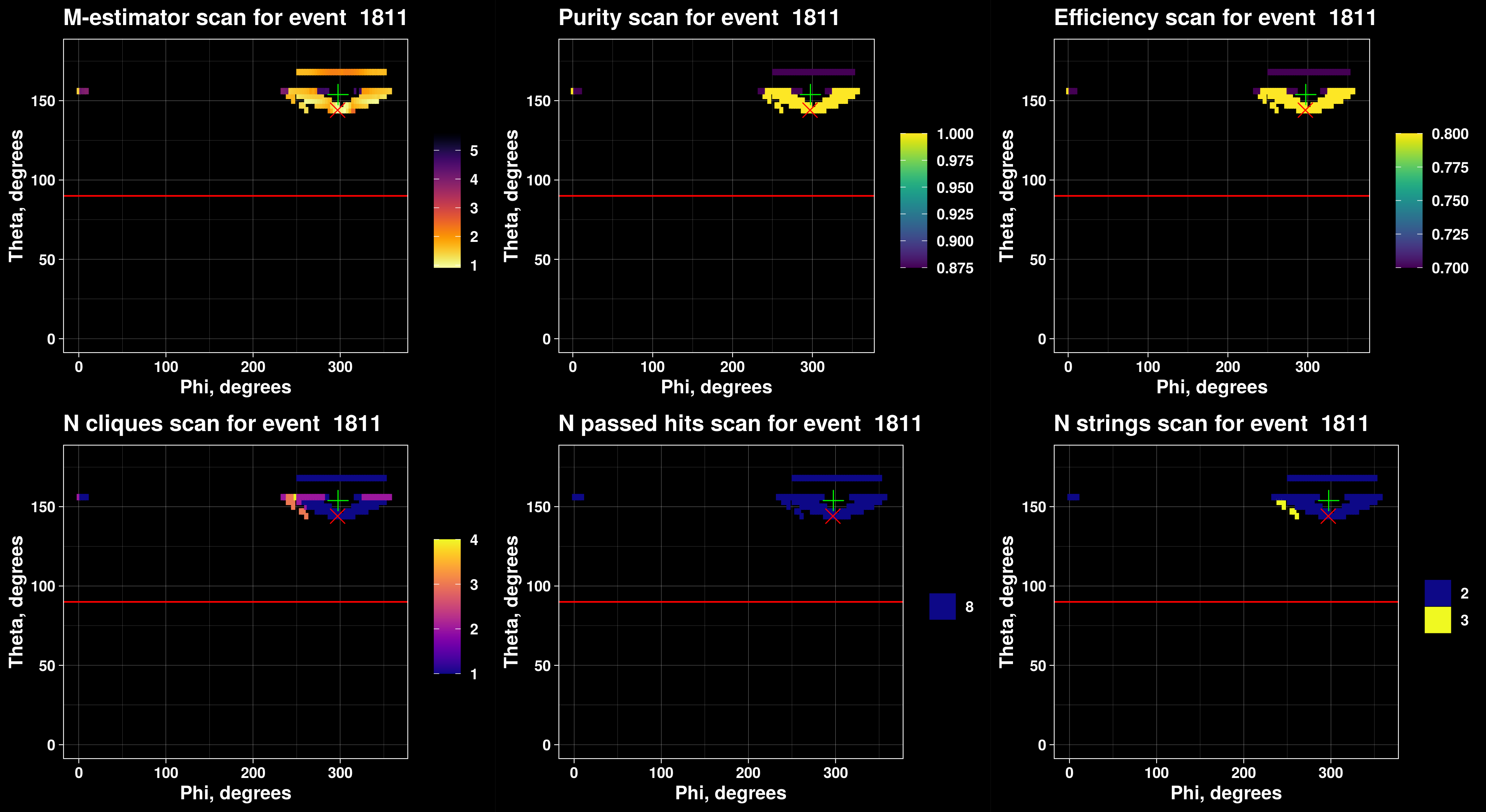}
         \caption{Downgoing atmospheric muon bundle}
         \label{fig:scan:down}
     \end{subfigure}
     \caption{Event scan examples. Each pixel corresponds to a scanned direction that passed the clique selection. Green cross corresponds to the true event origin, red cross to the selected direction.}
     \label{fig:scan_examples}
\end{figure}

\begin{figure}[t]
  \centering
  \includegraphics[height=3.5cm]{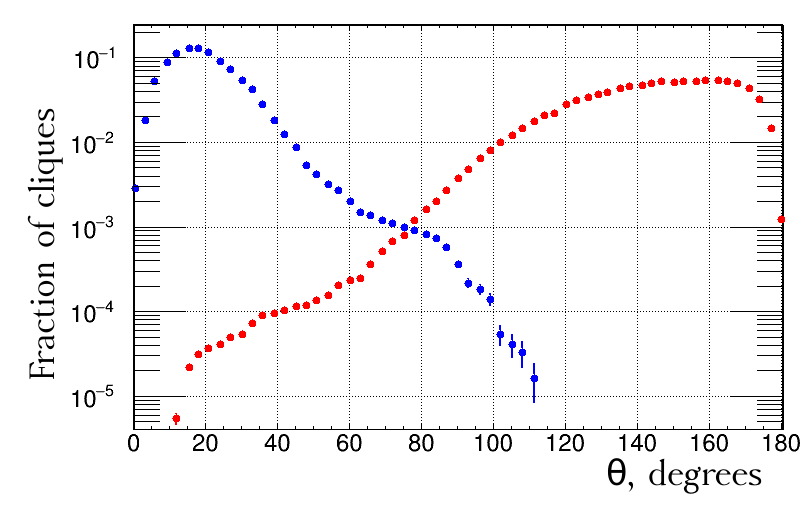}
  \caption{Distribution of cliques over the polar angle for an atmospheric neutrino (blue) and atmospheric muon bundle (red) samples.}
  \label{fig:thetas}
\end{figure}

The examples of a scan for two events are presented on Figure \ref{fig:scan_examples}.
Note that all the scan grid directions that passed clique selection are below the horizon for the upgoing event and above the horizon for the downgoing event.
As can be seen on Figure \ref{fig:thetas}, this remains the case for most of the vertices in the case of upgoing muons from atmospheric neutrinos and downgoing bundles, allowing for improved rejection of the atmospheric muon background in further analysis.

\section{Conclusion}
We have developed an efficient hit selection technique for suppressing noise hits in Baikal-GVD based on the approach presented in \cite{bruijn2008antares}.
Our approach alters key steps in the previous work by considering pairwise causal relationships between PMT hits as an undirected graph and assuming that signal hits should form fully connected subgraphs (cliques) in it.
The new approach has demonstrated a 95\% purity and 95\% efficiency for an atmospheric neutrino sample, showing almost two-fold inrease in efficiency compared to the hit selection algorithm previously used in Baikal-GVD.
The algorithm has shown comparable results for multicluster and high energy neutrino samples (see Table \ref{tbl:results}).
Besides noise suppression, this approach yields useful information about event configuration which can be used further along the processing pipeline.
In Baikal-GVD this approach is used in the latest version of the muon reconstruction, as described in \cite{reco2021}.

\section{Acknowledgements}
The work was partially supported by RFBR grant 20-02-00400.
The CTU group acknowledges the support by European Regional Development Fund-Project No. CZ.02.1.01/0.0/0.0/16\_019/0000766.
We also acknowledge the technical support of JINR staff for the computing facilities (JINR cloud).

\bibliographystyle{unsrt}
\bibliography{procbib}

\begin{thebibliography}{1}

\bibitem{gvd}
\url{https://baikalgvd.jinr.ru}.

\bibitem{gvd2021}
V.~A. Allakhverdyan et~al.
\newblock Neutrino telescope in lake {B}aikal: Present and nearest future.
\newblock {\em {\em PoS} {\bfseries ICRC2021} (these proceedings) 002}.

\bibitem{noise2019icrc}
A.~D. Avrorin et~al.
\newblock {The optical noise monitoring systems of the Lake {B}aikal
  environment for the {B}aikal-{GVD} telescope}.
\newblock {\em PoS}, ICRC2019:875, 2021.

\bibitem{lum2021}
V.~A. Allakhverdyan et~al.
\newblock The {B}aikal-{GVD} neutrino telescope as an instrument for studying
  {B}aikal water luminescence.
\newblock {\em {\em PoS} {\bfseries ICRC2021} (these proceedings) 400}.

\bibitem{bruijn2008antares}
Ronald Bruijn et~al.
\newblock {\em The Antares Neutrino Telescope: Performance studies and analysis
  of first data}.
\newblock Universiteit van Amsterdam, 2008.

\bibitem{allakhverdyan2021measuring}
V.~A. Allakhverdyan et~al.
\newblock Measuring muon tracks in {B}aikal-{GVD} using a fast reconstruction
  algorithm.
\newblock {\em arXiv preprint arXiv:2106.06288}, 2021.

\bibitem{bron1973algorithm}
Coen Bron and Joep Kerbosch.
\newblock Algorithm 457: finding all cliques of an undirected graph.
\newblock {\em Communications of the ACM}, 16(9):575--577, 1973.

\bibitem{reco2021}
V.~A. Allakhverdyan et~al.
\newblock Performance of the muon track reconstruction with the {B}aikal-{GVD}
  neutrino telescope.
\newblock {\em {\em PoS} {\bfseries ICRC2021} (these proceedings) 1080}.

\end{thebibliography}

\end{document}